\pdfoutput=1

\documentclass[11pt]{article}

\usepackage[]{acl}

\usepackage{times}
\usepackage{latexsym}
\usepackage{amsmath,amsfonts,amssymb}

\usepackage[T1]{fontenc}

\usepackage[utf8]{inputenc}

\usepackage{microtype}
\usepackage{algorithm}
\usepackage{algorithmic}
\usepackage[utf8]{inputenc} 
\usepackage[T1]{fontenc}    
\usepackage{url}            
\usepackage{booktabs}       
\usepackage{amsfonts}       
\usepackage{nicefrac}       
\usepackage{microtype}      
\usepackage{xcolor}         
\usepackage{enumitem}
\usepackage{multirow}
\usepackage{graphicx}
\usepackage{amsmath}
\usepackage{multirow}

\title{A Study on the Efficiency and Generalization of Light Hybrid Retrievers}

\author{
    Man Luo \textsuperscript{\rm 1}\footnotemark[1]
    \qquad Shashank Jain\textsuperscript{\rm 2}
    \qquad Anchit Gupta\textsuperscript{\rm 2}\footnotemark[2]
    \qquad Arash Einolghozati\textsuperscript{\rm 2}\footnotemark[2]
    \\
    \textbf{
    \qquad Barlas Oguz\textsuperscript{\rm 2}\footnotemark[2]
    \qquad Debojeet Chatterjee\textsuperscript{\rm 2}\footnotemark[2]
    \qquad Xilun Chen\textsuperscript{\rm 2}\footnotemark[2]
    \qquad Chitta Baral\textsuperscript{\rm 1}
    \qquad Peyman Heidari \textsuperscript{\rm 2}} \\
    \textsuperscript{\rm 1} Arizona State University \\
    \textsuperscript{\rm 2} Meta Reality Lab  \\
    \textsuperscript{\rm 1} \texttt{\{mluo26, chitta\}@asu.edu}\\
    \textsuperscript{\rm 2}\texttt{\{shajain, anchit, arashe, barlaso, debo, xilun, peymanheidari\}@fb.com}
}

\begin{document}
\maketitle
\begin{abstract}




Hybrid retrievers can take advantage of both sparse and dense retrievers. 
Previous hybrid retrievers leverage indexing-heavy dense retrievers. In this work, we study \textit{``Is it possible to reduce the indexing memory of hybrid retrievers without sacrificing performance?''} 
Driven by this question, we leverage an indexing-efficient dense retriever (i.e. DrBoost) and introduce a LITE retriever that further reduces the memory of DrBoost.
LITE is jointly trained on contrastive learning and knowledge distillation from DrBoost. 
Then, we integrate BM25, a sparse retriever, with either LITE or DrBoost to form light hybrid retrievers. 
Our Hybrid-LITE retriever saves $13\times$ memory while maintaining $98.0\%$ performance of the hybrid retriever of BM25 and DPR. 
In addition, we study the generalization capacity of our light hybrid retrievers on out-of-domain dataset and a set of adversarial attacks datasets.
Experiments showcase that light hybrid retrievers achieve better generalization performance than individual sparse and dense retrievers. 
Nevertheless, our analysis shows that there is a large room to improve the robustness of retrievers, suggesting a new research direction. 


 

\end{abstract}

\section{Introduction}

The classical IR methods, such as BM25~\cite{robertson2009probabilistic}, produce sparse vectors for question and documents based on bag-of-words approaches. 
Recent research pays attention toward building neural retrievers which learn dense embeddings of the query and document into a semantic space~\cite{karpukhin2020dense,khattab2020colbert}.
Sparse and dense retrievers have their pros and cons, and the hybrid of sparse and dense retrievers can take advantage of both worlds and achieve better performance than individual sparse and dense retrievers. 
Therefore, hybrid retrievers are widely used in practice~\cite{ma2021replication,chen2021salient}.

\begin{figure}
    \centering
    \includegraphics[width=\linewidth]{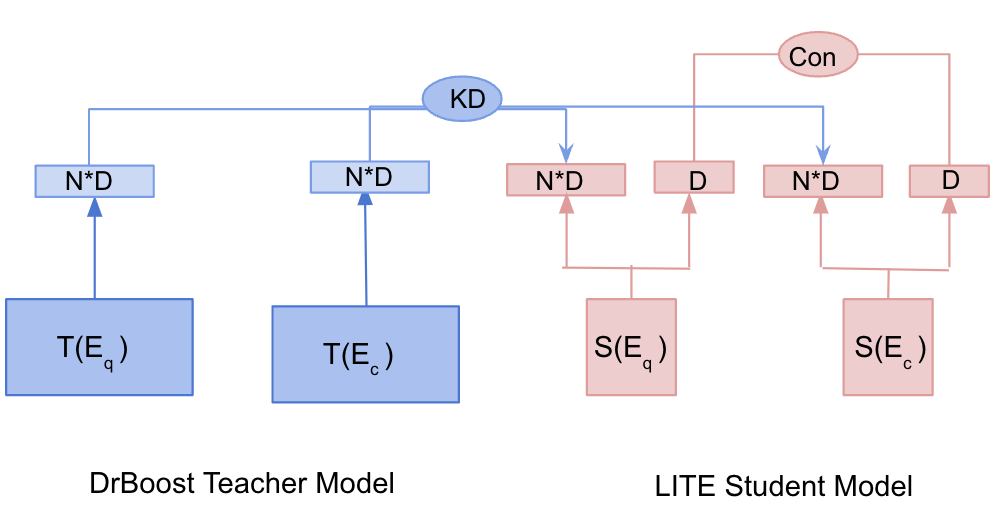}
    \caption{The teacher model (DrBoost) consists of N weak-learners and produces embeddings of dimension N*D. The student model (LITE) has one weak-learner and produces two embeddings: one has dimension of D, and one has dimension of N*D. The smaller embeddings learn to maximize the similarity between question and positive context embeddings, and the larger embeddings learn the embeddings from the teacher model.}
    \label{fig:teaser}
\end{figure}
Previous hybrid retrievers are composed of indexing-heavy dense retrievers (DR), in this work, we study the question \textit{``Is it possible to reduce the indexing memory of hybrid retrievers without sacrificing performance?''} 
To answer this question, we reduce the memory by using the state-of-the-art indexing-efficient retriever, DrBoost~\cite{lewis2021boosted}, a boosting retriever with multiple ``weak'' learners.
Compared to DPR~\cite{karpukhin2020dense}, a representative DR, DrBoost reduces the indexing memory by 6 times while maintaining the performance.
We introduce a LITE model that further reduces the memory of DrBoost, which is jointly trained on retrieval task via contrastive learning and knowledge distillation from DrBoost (see Figure \ref{fig:teaser}). 
We then integrate BM25 with either LITE and DrBoost to form light hybrid retrievers (Hybrid-LITE and Hybrid-DrBoost) to assess whether light hybrid retrievers can achieve memory-efficiency and sufficient performance.  

We conduct experiments on the NaturalQuestion dataset~\cite{kwiatkowski2019nq} and draw interesting results. First of all, LITE retriever maintains $98.7\%$ of the teacher model performance and reduces its memory by $2$ times. 
Second, our Hybrid-LITE saves more than $13\times$ memory compared to Hybrid-DPR, while maintaining more than $98.0\%$ performance; and Hybrid-DrBoost reduces the indexing memory ($8\times$) compared to Hybrid-DPR and maintains at least $98.5\%$ of the performance. 
This shows that the light hybrid model can achieve sufficient performance while reducing the indexing memory significantly, which suggests the practical usage of light retrievers for memory-limited applications, such as on-devices. 

One important reason for using hybrid retrievers in real-world applications is the generalization. Thus, we further study if reducing the indexing memory will hamper the generalization of light hybrid retrievers. Two prominent ideas have emerged to test generalization: 
out-of-domain (OOD) generalization and adversarial robustness~\cite{gokhale2022generalized}.
We study OOD generalization of  retrievers on EntityQuestion~\cite{sciavolino2021simple}.
To study the robustness, we leverage six techniques~\cite{morris2020textattack} to create adversarial attack testing sets based on NQ dataset. Our experiments demonstrate that Hybrid-LITE and Hybrid-DrBoost achieve better generalization performance than individual components. 
The study of robustness shows that hybrid retrievers are always better than sparse and dense retrievers. 
Nevertheless all retrievers are vulnerable, suggesting room for improving the robustness of retrievers, and our datasets can aid the future research.

\section{Related Work}
\paragraph{Hybrid Retriever}
integrates the sparse and dense retriever and ranks the documents by interpolating the relevance score from each retriever.
The most popular way to obtain the hybrid ranking is applying linear combination of the sparse/dense retriever scores~\cite{karpukhin2020dense,ma2020zero,luan2021sparse,ma2021simple,luo2022improving}.
Instead of using the scores, \citet{chen2022out} adopts Reciprocal Rank Fusion~\cite{cormack2009reciprocal} to obtain the final ranking by the ranking positions of each candidate retrieved by individual retriever.
\citet{arabzadeh2021predicting} trains a classification model to select one of the retrieval strategies: sparse, dense or hybrid model.
Most of the hybrid models rely on heavy dense retrievers, and one exception is \cite{ma2021simple}, where they use linear projection, PCA, and product quantization~\cite{jegou2010product} to compress the dense retriever component. 
Our hybrid retrievers use either DrBoost or our proposed LITE as the dense retrievers, which are more memory-efficient and achieve better performance than the methods used in \cite{ma2021simple}.

\paragraph{Indexing-Efficient Dense Retriever.}
Efficiency includes two dimensions: latency~\cite{seo2019real,lee2021learning,varshney2022can} and memory. 
In this work, our primary focus is on memory, specifically the memory used for indexing.
Most of the existing DRs are indexing heavy~\cite{karpukhin2020dense,khattab2020colbert,luo2022neural}.
To improve the indexing efficiency, there are mainly three types of techniques. 
One is to use vector product quantization~\cite{jegou2010product}.
Second is to compress a high dimension dense vector to a low dimension dense vector, for e.g. from 768 to 32 dimension~\cite{lewis2021boosted,ma2021simple}.  
The third  way is to use a binary vector~\cite{yamada2021efficient,zhan2021jointly}. 
Our proposed method LITE (\S\ref{sec:kd}) reduces the indexing memory by joint training of retrieval task and knowledge distillation from a teacher model. 

\paragraph{Generalization of IR.}
Two main benchmarks have been proposed to study the OOD generalization of retrievers, BEIR~\cite{thakur2021beir} and EntityQuestion~\cite{sciavolino2021simple}.
As shown by previous work~\cite{thakur2021beir,chen2022out}, the generalization is one major concern of DR. 
To address this limitation, \citet{wang2021gpl} proposed GPL, a domain adaptation technique to generate synthetic question-answer pairs in specific domains. 
A follow-up work~\citet{thakur2022domain} trains BPR and JPQ  on the GPL synthetic data to achieve efficiency and generalization. 
~\citet{chen2022out} investigates a hybrid model in the OOD setting, yet different from us, they use a heavy DR and do not concern the indexing memory. 
Most existing work studies OOD generalization, and much less attention paid toward the robustness of retrievers~\cite{penha2022evaluating,zhuang2022characterbert,chentowards}. 
To study robustness, \citet{penha2022evaluating} identifies four ways to change the syntax of the queries but not the semantics. 
Our work is a complementary to \citet{penha2022evaluating}, where we leverage adversarial attack techniques~\cite{morris2020textattack} to create six different testing sets for NQ dataset~\cite{kwiatkowski2019nq}.



\section{Model}
In this section, we first review DrBoost~\cite{lewis2021boosted}, and our model LITE which further reduces the memory of DrBoost, and lastly, we describe the hybrid retrievers that integrate light dense retrievers (i.e. LITE and DrBoost) and BM25. 

\subsection{Reivew of DrBoost} 
DrBoost is based on ensemble learning to form a strong learner by a sequence of weak leaners, and each weak learner is trained to minimize the mistakes of the combination of the previous learners. 
The weak learner has the similar architecture as DPR~\cite{karpukhin2020dense} (review of DPR is given in Appendix~\ref{apd:preliminary}), but the output vectors are compressed to a much lower dimension by a linear regression layer $\mathrm{W}$, 
\begin{equation}
\mathrm{v}_q^{i} = \mathrm{W}_{q} \cdot \mathrm{V}_q^{i}, \quad \mathrm{v}_c^{i} = \mathrm{W}_{c} \cdot  \mathrm{V}_c^{i},
    \label{eq:drboost_enc1}
\end{equation}
where $\mathrm{V}_{q/c}^{i}$ are the representation of question/document given by the embeddings of special tokens \texttt{[CLS]} of a high dimension, $\mathrm{v}_{q/c}^{i}$ are the lower embeddings produced by the $i^{th}$ weak learner.
The final output representation of DrBoost is the concatenation of each weak learners' representations as expressed by Eq. \ref{eq:drboost_enc2}. 

\begin{equation}
\boldsymbol{\mathrm{q}} = [\mathrm{v}_q^{1}, \dots, \mathrm{v}_q^{n}], \quad \boldsymbol{\mathrm{c}} = [\mathrm{v}_c^{1}, \dots, \mathrm{v}_c^{n}],
\label{eq:drboost_enc2}
\end{equation}
where $n$ is the total number of weak learners in the DrBoost. 
The training objective of DrBoost is
\begin{equation}
    \mathcal{L}_{con} = -\log \frac{e^{\mathrm{sim}(q, c^{+})}}{e^{\mathrm{sim}(q, c^{+})}+ \sum_{j=1}^{j=n}e^{\mathrm{sim}(q, c_{j}^{-})}},
    \label{eq:dpr_loss1}
\end{equation}
where $\mathrm{sim}(q, c)$ is the inner-dot product.

\subsection{LITE: Joint Training with Knowledge Distillation} \label{sec:kd}
Since DrBoost has N encoders, the computation of query representations takes N times as a single encoder. 
To save latency, \citet{lewis2021boosted} trains a student encoder which learns the N embeddings from the teacher encoders.
As a result, while the student model consists of only one encoder, it produces the same indexing memory as the teacher model.  
Here,  we want to further reduce the student indexing memory.
To achieve this,  we introduce a LITE retriever (see Figure \ref{fig:teaser}), which produces two embeddings for an input text: one has a smaller dimension ($\mathrm{v}_{q/c, s}$) for retrieval task, and the other one is a larger dimension ($\mathrm{v}_{q/c, l}$) for learning knowledge from the N teacher models.   
The small and large embeddings are obtained by compressing the \texttt{[CLS]} token embedding via separate linear regression layers, mathematically, 
\begin{equation}
\mathrm{v}_{q/c, s} = \mathrm{W}_{q/c, s} \cdot \mathrm{V}_{q/c},  \quad  \mathrm{v}_{q/c, l} = \mathrm{W}_{q/c, l} \cdot \mathrm{V}_{q/c}
    \label{eq:distill}
\end{equation}
$\mathrm{v}_{q/c, s}$ is optimized by the contrastive loss (E.q. \ref{eq:dpr_loss1}). 
And $\mathrm{v}_{q/c, l}$ learns the teacher model embeddings. The knowledge distillation (KD) loss is composed of three parts (Eq.~\ref{eq:kd_loss}): 1) the distance between student question embeddings and the teacher question embeddings, 2)  the distance between student context embeddings and the teacher context embeddings, and 3) the distance between student question embeddings and the teacher positive context embeddings. 
\begin{equation}
\mathcal{L}_{KD} = \lVert \mathrm{v}_{q, l} - \boldsymbol{\mathrm{q}} \rVert^{2}  + \lVert \mathrm{v}_{c, l}- \boldsymbol{\mathrm{c}} \rVert^{2} + \lVert \mathrm{v}_{q, l}- \boldsymbol{\mathrm{c}}^{+} \rVert^{2}
\label{eq:kd_loss}
\end{equation}
The final objective of the student model is,
\begin{equation}
\mathcal{L}_{joint}  = \mathcal{L}_{con} +\mathcal{L}_{KD}.
\label{eq:joint_loss}
\end{equation}

In contrast to the distillation method in DrBoost, which solely learns the embeddings from the teacher model, LITE is simultaneously trained on both the retrieval task and the knowledge distillation task. During the inference time, LITE only utilizes the retrieval embeddings ($\mathrm{v}_{c,s}$ ) to achieve indexing-efficiency. 
It is also notable that LITE is a flexible training framework capable of incorporating most neural retrievers as its backbone models, despite our work being solely reliant on DrBoost.

\subsection{Memory Efficient Hybrid Model} \label{sec:hybrid}
Our hybrid models retrieve the final documents in a re-ranking manner. 
We first retrieve the top-k documents using BM25 and dense retriever (DrBoost or LITE) separately.
The document scores produced by these two retrievers are denoted by $S_\mathrm{BM25}$ and $S_\mathrm{DR}$ respectively.
We apply  MinMax normalization to original socres to obtain $S_{BM25}^\prime$ and $S_{DR}^\prime$ ranging from $[0, 1]$. 
For each document, we get a new score for final ranking:
\begin{equation}
    S_\mathrm{hybrid} = w_{1} \times S_\mathrm{BM25}^\prime+  w_{2} \times S_\mathrm{DR}^\prime, 
    \label{eq:hybrid_score}
\end{equation}
where $w_1$ and $w_2$ denote the weights of BM25 and DrBoost scores respectively. 
In our experiments, we simply set equal weights (i.e. 0.5) to each method. 
If a context is not retrieved by either retriever, then its score for that retriever is $0$.  

\section{Adversarial Attack Robustness Dataset}
\begin{figure}
    \centering
    \includegraphics[width=\linewidth]{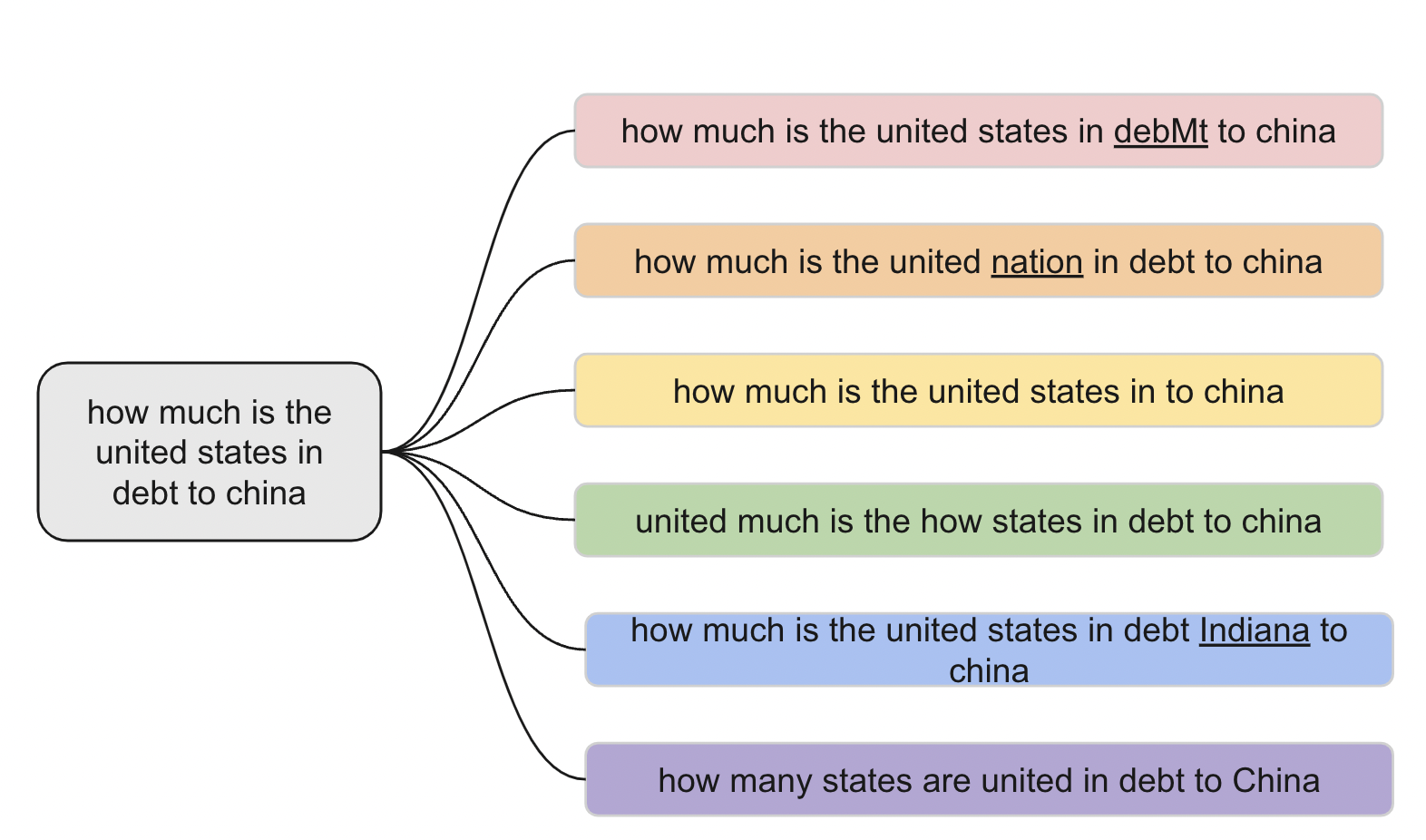}
    \caption{Examples of the adversarial attack questions. Underline denotes the change from the original question. The examples from the top to the bottom are augmented by CS, WD, SR, WOS, SI, and BT.}
    \label{fig:example_rob}
\end{figure}

Adversarial attacks are used to asses model's robustness, 
where testing samples are obtained by small perturbations of the original samples, and such perturbations keep the  label unchanged. 
To test the robustness of IR systems, we create 6 different adversarial attacks\footnote{We use TextAttack library~\cite{morris2020textattack}.}  for NQ~\cite{kwiatkowski2019nq}.
Each method is chosen because they do not change the original meaning of the queries and the relevant documents should be the same as the original relevant documents (see Figure  \ref{fig:example_rob}).
The six methods include: \textit{Char-Swap (CS):} augments words by swapping characters out for other characters;
\textit{Word Deletion (WD):} delete a word randomly from the original query;
\textit {Synonym Replacement (SR):} replaces a word in the query with a synonym from the WordNet~\cite{miller1995wordnet};
\textit {Word-Order-Swap (WOS):} swaps the order of the words in the original query; 
\textit {Synonym Insertion (SI):} insert a synonym of a word from the WordNet to the original query;
\textit {Back-Translation (BT)} translates the original query into a target language and translates it back to the source language. 
Figure \ref{fig:example_rob} shows an example of each attacked instance\footnote{The adversarial robustness dataset is available in \href{https://github.com/facebookresearch/dpr-scale}{this link}.}.

\section{Experiments and Results}
\paragraph{Existing Methods.}
We include four existing methods in this work, DrBoost~\cite{lewis2021boosted}, DPR~\cite{karpukhin2020dense}, SPAR~\cite{chen2021salient} and a heavy hybrid model BM25 + DPR~\cite{karpukhin2020dense}. 
In Table \ref{tab:nq_eq}, the performance of DrBoost is from the original paper and the performance of the other three methods are from~\cite{chen2021salient}. 

\paragraph{Our Baselines.} 
Three baselines are presented, BM25, DPR$_{32}$, and DrBoost-2.
DPR$_{32}$ refers to DPR with a linear projection layer to representation to 32 dimension. 
DrBoost-2 takes DPR$_{32}$ as the first weak learner, and uses it to mine negative passages to train the next weak learner
and then combine these two models.
We do not go beyond 2 weak learners because our goal is to achieve memory-efficiency while increasing the number of encoders in the DrBoost will yield larger indexing. 

\paragraph{Our Models.} LITE and the three light hybrid models are presented. 
LITE is trained by the method we introduce in \S\ref{sec:kd} with the distilled knowledge from DrBoost-2 teacher model. 
We present three hybrid models BM25 + LITE, BM25 + DPR$_{32}$, and BM25 + DrBoost-2, which are memory-efficient compared to existing methods.  
Next we present the experiments and the findings. 


\subsection{Memory Efficiency and Performance}

\paragraph{LITE} achieves  much better performance compared to DPR$_{32}$ even though both use the same amount of memory. LITE also maintains more than $98\%$ knowledge of its teacher (DrBoost-2), and importantly saves $2\times$  of indexing memory. 
Such results shows the effectiveness of LITE. 

\paragraph{Hybrid-LITE} achieves better performance than DrBoost-2 while using less indexing memory. Hybrid-LITE also matches the performance of DrBoost in terms of R@100 (87.4 v.s. 87.2) while using  $3\times$ less memory. 
Compared with Hybrid-DPR, Hybrid-LITE maintains 98.4\% performance but uses $13\times$ less memory.  
Compared with the SOTA model SPAR, Hybrid-LITE achieves 98.2\% performance and uses $25\times$ less memory. 

\paragraph{Hybrid-DrBoost-2}  achieves almost similar performance as DrBoost which contains 6 encoders.
This shows the effects of BM25 match the capacity of 4 encoders in the DrBoost. 
We also compare Hybrid-DrBoost-2 with BM25 + DRP or SPAR, where our model achieves almost 99\% performance but uses less than 8$\times$ or 16$\times$ of memory.

\begin{table}[]
    \centering
    \resizebox{\linewidth}{!}{
    \begin{tabular}{lccccc}
    \toprule
    \multirow{2}{*}{Method} & \multirow{2}{*}{Index-M } &  \multicolumn{2}{c}{NQ} & \multicolumn{2}{c}{EntityQuestion}\\
     \cmidrule(lr){3-4}\cmidrule(lr){5-6}
    & (GB) & R@20 & R@100 & R@20 & R@100\\
    \toprule
    {\bf Existing Method} & ~ & ~ \\ 
    \hline
    DrBoost & 15.4/13.5 & 81.3 &  87.4 & 51.2 &  63.4 \\
    DPR  & 61.5 &  79.5 & 86.1 & 56.6 & 70.1 \\
    BPR & 2 & 77.9& 85.7 & - & - \\
    BM25+DPR & 63.9 & 82.6 & 88.6 & 73.3 & 82.3 \\
    SPAR & 123.0 &  {83.6}& {88.8} & {74.0} & {82.0}\\
    
    \hline
    {\bf Our Baseline} & ~ & ~ & ~\\
    \hline
    BM25 & 2.4 & 63.9 & 78.8 & 71.2 & 79.7 \\
    DPR$_{32}$ & 2.5 & 70.4 & 80.0 & 31.1 & 45.5 \\
    DrBoost-2  & 5.1 & 77.3 & 84.5 & 41.3 & 54.2 \\
    \hline
    {\bf Our Model } & ~ &~ & ~\\
    \hline
    LITE & 2.5 & 75.1 & 83.4  & 35.0 & 48.1 \\
    Hybrid-LITE & 4.9 & 79.9 & 87.2 & 71.5 & 80.8 \\
    Hybrid-DPR$_{32}$  & 4.9 & 77.7 & 86.2 & 70.8 & 80.5 \\
    Hybrid-DrBoost-2  & 7.5 & 80.4 & 87.5 & 72.4 & 81.4\\
    \bottomrule
    \end{tabular}
    }
    \caption{Performance of existing methods, our baselines and our hybrid model on NQ dataset. The performance of DrBoost on NQ is using 6 weak learners (15.4 GB indexing memory) and of EntityQuestion is using 5 weak learners (13.5 GB).}
    \label{tab:nq_eq}
\end{table}

\subsection{Out-of-Domain Generalization}
We study the out-of-domain generalization of retriever on 
EntityQuestion~\cite{sciavolino2021simple}, which consists of simple entity centric questions but shown to be difficult for dense retrievers.
We train the model on NQ and test on EQ.
 
First of all, our experimental results show  that the performance of DPR$_{32}$, DrBoost-2, and LITE are much worse than BM25 on EQ. 
Nevertheless, our hybrid models improve both BM25 and dense retriever performance.
Our light hybrid models achieve similar performance as hybrid-DPR and SPAR, which demonstrates that our light hybrid retrievers exhibit good OOD generalization. 





\subsection{Adversarial Attack Robustness}

The robustness is evaluated in terms of both performance (higher R@K means more robust) and the average drop w.r.t the original performance on NQ dataset (smaller drop means more robust). 

From Table \ref{tab:robustness}, we observe that all models perform worse compared to the original performance on all adversarial attack sets, which showcase that the current retrievers are not robust enough. 
Interestingly, 
while it is expected that BM25 will be robust on word-order-swap (WOS) attack, it is not straightforward that a dense retriever is also robust on this type of questions.
This shows that the order of the words in the question is not important for the dense retriever neither. 
We also see that char-swap (CS) is the most difficult attack, which means that both types of retrievers might not perform well when there are typos in the questions. 

Diving into the individual performance of each retriever, we see that some models are more robust than others. 
For example, LITE is more robust than DPR$_{32}$. 
We also compare the hybrid model with the pure dense retriever counterparts (e.g. compare hybrid Drboost-2 with DrBoost-2), and find that hybrid models are consistently more robust. 
This suggests that the hybrid model can mitigate the performance drop of both BM25 and dense retriever.

\begin{table}[]
    \centering
    \resizebox{\linewidth}{!}{
    \begin{tabular}{lccccccccc}
    \toprule
    \multirow{2}{*}{Method}  & \multicolumn{8}{c}{R@100} \\
     \cmidrule(lr){2-9}
    & Ori & CS & WD & SR & WOS & SI & BT & Drop \\
    \toprule
    BM25    & 78.8 & 68.2 & 71.7 & 74.5 & 78.3 & 77.2 & 71.2 & 5.9 \\
    DPR$_{32}$ & 80.8 & 61.9 & 65.8 & 75.3 & 76.4 & 73.3 & 71.1 & 10.3 \\
    LITE & 83.4 & 69.3 & 71.8 & 78.9 & 81.2 & 79.0 & 75.6 & 7.9 \\
    DrBoost-2 & 84.5 & 71.6 & 80.1 & 74.7 & 82.6 & 80.4 & 77.9 & 7.8 \\
    DPR$_{768}$ & 86.1 & 74.8 & 78.9 & 82.5 & 85.0 & 83.4 & 80.3 & 5.5 \\
    \midrule
    +DPR$_{32}$ & 86.2 & 74.4 & 78.0 & 82.7 & 84.9  & 83.2 & 78.6 & 6.1 \\
    +LITE  &  87.2 & 76.5 & 78.0 & 83.7 & 86.6 & 85.4 & 80.8 & 5.1 \\
    +DrBoost-2 & 87.5 & 77.7 & \textbf{84.6} & 81.0 & 86.7 & 85.9 & 81.9  & 5.2 \\
    +DPR$_{768}$ & \textbf{88.3} & \textbf{78.6} & 82.9 & \textbf{85.4} & \textbf{87.7} & \textbf{86.6} & \textbf{82.6}  & \textbf{4.4} \\
    \bottomrule
    \end{tabular}
    }
    \caption{Ori: Original question; CS: CharSwap; WD:Word deletion; WSR: WordNet synonym replacement; WOR: Word order swaps; RSI :Random synonym insertion; BT: Back Translation. The smaller the Average Drop is, the more robust the model is.}
    \label{tab:robustness}
\end{table}


\section{Conclusion}
To achieve indexing efficiency, in this work, we study light hybrid retrievers. We introduce LITE, which is jointly trained on retrieval task via contrastive learning and knowledge distillation from a more capable teacher models which requires heavier indexing-memory. 
While in this work, we mainly take DrBoost as the teacher model, LITE is a flexible training framework that can be incorporated with most of the neural retriever. 
Then, we integrate  BM25 with LITE or DrBoost to form light hybrid retrievers.
Our light hybrid models achieve sufficient performance and largely reduce the memory. 
We also study the generalization of retrievers and suggest that all sparse, dense, and hybrid retrievers are not robust enough, which opens up a new avenue for research.

\section*{Limitation}
The main limitation of this work is the technical novelty of hybrid retriever. 
Hyrbid-DrBoost is built on top of DrBoost, and  the interpolation of BM25 with DrBoost. 
However, we would like to point out that our study can serve as an important finding for real-life applications. Previous retrievers are built on top of indexing-heavy dense retrievers, such as DPR. This limits their applications where memory is a hard constraints, for example, on-devices. 
Our study suggests that a light hybrid retriever can save memory but maintain sufficient performance.


\bibliography{anthology,custom}

\appendix
\newpage
\section{Preliminary} \label{apd:preliminary}
\paragraph{BM25}
\citet{robertson2009probabilistic}, is a bag-of-words ranking function that scores the query (Q) and document (D) based on the term frequency. 
The following equation is the one of the most prominent instantiations of the function, 
\begin{equation}
\begin{split}
score(D, Q) = \sum_{i=1}^{n} \mathrm{IDF}(q_i) \cdot \\
 \frac{f(q_i, D) \cdot (k_1+1)}{f(q_i, D)+k1\cdot(1-b+b\cdot \frac{|D|}{avgdl})},
\end{split}
\end{equation}
where $\mathrm{IDF}(q_i)$ is the inverse document frequency of query term $q_i$, $f(q_i, D)$ is the frequency of $q_i$ in document $D$, $|D|$ is the length of the document $D$, and $avgdl$ is the  average length of all documents in the corpus. 
In practice, $k_1 \in [1.2, 2.0]$ and $b=0.75$.
BM25 is an unsupervised method that generalizes well in different domains~\cite{Thakur2021BEIRAH}.

\paragraph{DPR} 
Dense passage retriever involves two encoders: the question encoder $\mathrm{E}_q$ produces a dense vector representation $\mathrm{V}_q$ for an input question $q$, and the context encoder $\mathrm{E}_c$ produces a dense vector $\mathrm{V}_c$ representation for an input context $c$. 
Both encoders are BERT models and the output vectors  are the embeddings of the special token \texttt{[CLS]} in front of the input text (Eq. \ref{eq:dpr_enc}).
\begin{equation}
    \mathrm{V}_q = \mathrm{E}_q(q)\texttt{[CLS]}, \quad \mathrm{V}_c = \mathrm{E}_c(c)\texttt{[CLS]}.
    \label{eq:dpr_enc}
\end{equation}

The score of $c$ w.r.t $q$ is the inner-dot product of their representations (Eq \ref{eq:dpr_sim}). 
\begin{equation}
    \mathrm{sim}(q, c) = \mathrm{V}_q^\top \mathrm{V}_c.
    \label{eq:dpr_sim}
\end{equation}
DPR uses contrastive loss to optimize the model such that the score of positive context $c^{+}$ is higher than the score of the negative context $c^{-}$. Mathematically, DPR maximizes the following objective function, 
\begin{equation}
    \mathcal{L}_{con} = -\log \frac{e^{\mathrm{sim}(q, c^{+})}}{e^{\mathrm{sim}(q, c^{+})}+ \sum_{j=1}^{j=n}e^{\mathrm{sim}(q, c_{j}^{-})}},
    \label{eq:dpr_loss}
\end{equation}
where $n$ is the number of negative contexts. 
For better representation learning, DPR uses BM25 to mine the hard negative context and the in-batch negative context to train the model.

\section{Ablation Study}

In this section, we conduct ablation studies to see the effects of the proposed methods, and all models are trained and tested on NQ dataset.

\subsection{LITE Can Improve DrBoost} 

Recall that DPR$_{32}$ is one encoder in DrBoost-2, and since LITE performs better than DPR$_{32}$ (see Table \ref{tab:nq_eq}), we ask the question can LITE replaces DPR$_{32}$ to form a stronger DrBoost-2 model? 
To answer this question, we compare the performance of R-DrBoost-2 (i.e. replace DPR$_{32}$ with LITE) with the original DrBoost-2.
From Table \ref{tab:kd_drboost}, 
We observe that R-DrBoost-2 performs worse than DrBoost-2, indicating that the encoders in the DrBoost indeed relate and complement to each other and replacing an unrelated encoder degrades the performance. 
Then we ask another question, can we train a weak learner that minimizes the error of LITE, and combine LITE with the new weak learner to form a stronger DrBoost (L-DrBoost-2)? 
Table \ref{tab:kd_drboost} shows L-DrBoost-2 is better than DrBoost-2, and hybrid L-DrBoost-2 is better than hybrid DrBoost-2 as well (81.0 v.s. 80.4 on R@20). 
This indicates that starting with a stronger weak learner can yield a stronger DrBoost. 

\begin{table}[]
    \centering
    \resizebox{\linewidth}{!}{
    \begin{tabular}{ccccc}
    \toprule
    Metric & O-DrBoost & R-DrBoost & LITE-DrBoost & H-LITE-DrBoost\\
    
    \toprule
    R@20 & 77.3 & 75.6 & 77.9 & \textbf{81.0} \\ 
    R@100 & 84.5 & 83.9 & 84.7 & \textbf{87.5}\\ 
    
    \bottomrule
    \end{tabular}
    }
    \caption{Three DrBoost (with 2 weak learners) and one hybrid retriever. O-DrBoost: the original DrBoost, R-DrBoost:replace the first weak learner in O-DrBoost with LITE, LITE-DrBoost: use LITE as the first weak learner and mine negative using LITE to train a new weak learner to form a DrBoost, H-LITE-DrBoost: hybrid BM25 with LITE-DrBoost.}
    \label{tab:kd_drboost}
\end{table}

\subsection{Hybrid model consistently improves the DrBoost performance.}
\begin{figure}
    \centering
    \includegraphics[width=\linewidth]{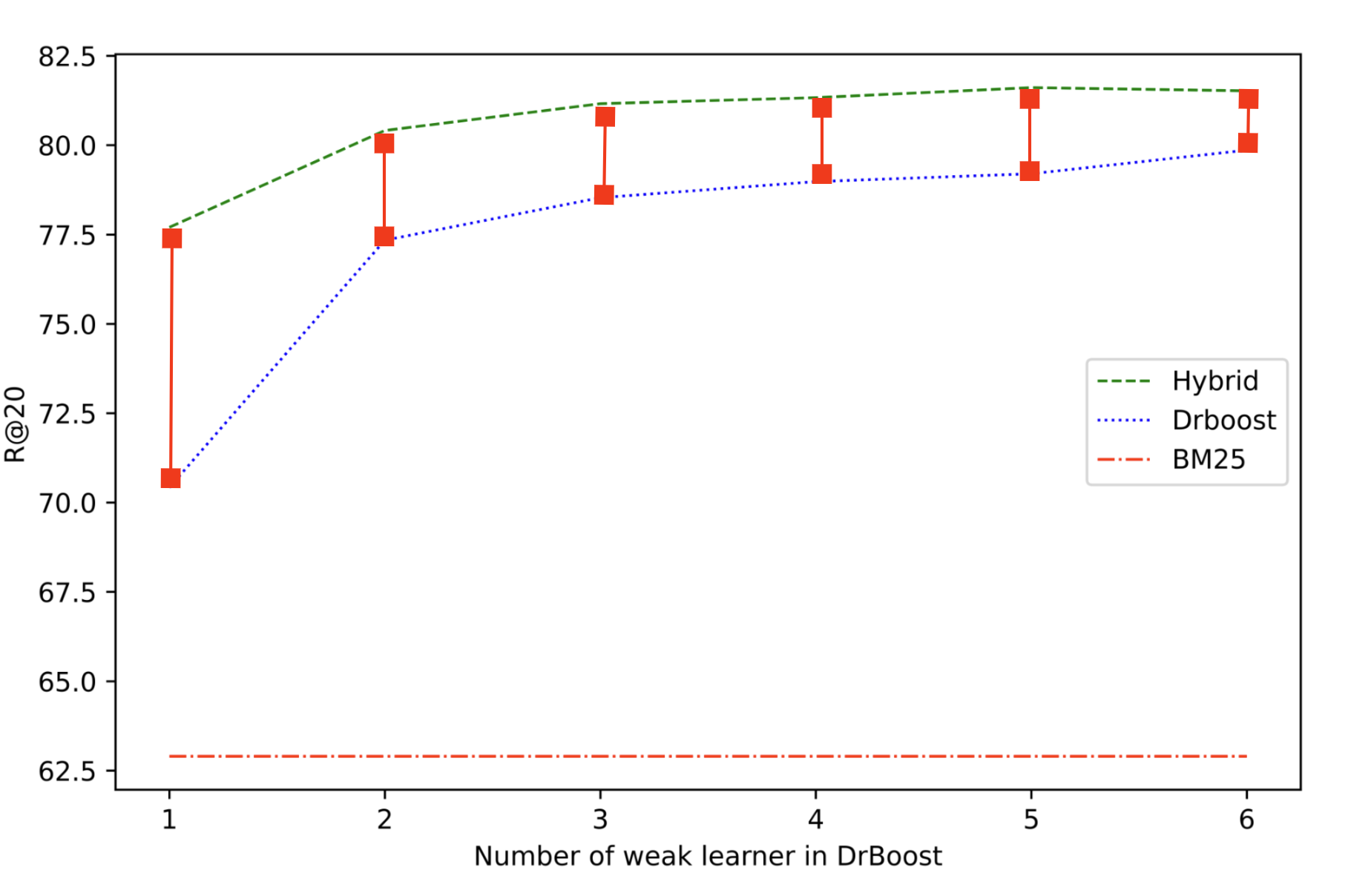}
    \caption{Compare DrBoost, BM25 and the Hybrid models performance.}
    \label{fig:drboost}
\end{figure}

We study six DrBoost models with 1-6 weak learners. 
In Figure \ref{fig:drboost}, we see that the performance of hybrid models consistently improves the DrBoost performance, demonstrating the results of BM25 and DrBoost complement each other and combining two models improves individual performance. 
We also see that the improvement is larger when the DrBoost is weaker, e.g. hybrid model significantly improves DPR$_{32}$. 

\begin{table}[]
    \centering
    \resizebox{\linewidth}{!}{
    \begin{tabular}{lccc}
    \toprule
    \multirow{2}{*}{Model}  & \multirow{2}{*}{Method} &  \multicolumn{2}{c}{NQ}\\
    \cmidrule(lr){3-4}
    & & R20 & R100\\
    \toprule
    \multirow{3}{*}{Hybrid(32*2)} 
      & Simple Sum & 79.03 & 84.63 \\
      & Multiplication & 79.03 & 84.63 \\
      & MinMax and Sum & \textbf{80.41} & \textbf{87.47} \\
    \midrule
    \multirow{3}{*}{Hybrid(32*6)} 
      & Simple Sum & \textbf{81.61} & 86.12 \\
      & Multiplication & 81.19 & 86.12 \\
      & MinMax and Sum & 81.52 & \textbf{88.28} \\
    \bottomrule
    \end{tabular}
    }
    \caption{Compare three hybrid scores. We study two hybrid model, BM25 with 2 weak learners (32*2) and BM25 with 6 weak learners (32*6)}
    \label{tab:diff_hybrid}
\end{table}

\subsection{Different Hybrid Scores}
In our hybrid model, besides the hybrid scores we introduced in \S\ref{sec:hybrid},  we also study two different hybrid scores of BM25 and the DrBoost. Simple Summation is to add two scores together, and multiplication is to mutiply two scores. 
We compare two hybrid models' performance, Hybrid-DrBoost-2 and Hybrid-DrBoost-6. 
Table \ref{tab:diff_hybrid} shows that the MinMax normalization performs the best (except that simple summation is slightly better in terms of R@20 for hybrid models with 6 weak learners).

\end{document}